\documentclass{moriond}

\usepackage{parskip}

\def\Journal#1#2#3#4{{#1} {\bf #2}, #3 (#4)}

\def\PRL{\em Phys. Rev. Lett.}
\def\PRD{{\em Phys. Rev.} D}

\def\PTEP{\em Prog. Theor. Exp. Phys.}

\def\be{\begin{equation}}
\def\ee{\end{equation}}
\def\bea{\begin{eqnarray}}
\def\eea{\end{eqnarray}}

\newcommand{\Photo}{}

\begin{document}
\vspace*{4cm}
\title{Weak decays of heavy-quark baryons}

\author{ Janina Nicolini\\On behalf of the LHCb collaboration }

\address{TU Dortmund University, Dortmund, Germany\\
Université Paris-Saclay, Orsay, France}

\maketitle\abstracts{
Weak decays of heavy-quark baryons offer an attractive laboratory to search for effects 
beyond the Standard Model (SM), complementary to searches in meson decays. However, the properties of these baryons are not well understood. In particular, the $\Omega^-_b$ baryon is the least 
studied weakly decaying $b$ baryon. Therefore, the most precise determination of the $\Omega^{-}_{b}$ 
baryon mass is reported, using $\Omega^-_b \to J/\psi \Omega^-$ decays at the LHCb experiment. 
In addition, the relative production rate of the $\Omega^-_b$ baryon in $pp$ collisions at the LHC is determined for the first time. These quantities are crucial to predict and measure branching fractions of $\Omega^{-}_{b}$ baryon decays.}

\section{Precision measurements of weakly decaying \texorpdfstring{$b$}{b} baryons}
Weak decays of heavy-quark baryons provide a complementary laboratory to search for effects
beyond the SM and offer rich angular structures compared to meson decays. These are 
particularly interesting as they allow to test spin dependencies of possible new physics 
scenarios. The quark model predicts four weakly decaying $b$ baryons with one heavy quark: the lightest one, the $\Lambda_b^0=|bud\rangle$, the isospin-doublet of $\Xi_b^0=|bsu\rangle$ and $\Xi_b^-=|bsd\rangle$
and the heaviest one, the $\Omega_b^-=|bss\rangle$. The experimental precision of the measured $b$ baryon properties scales largely with the production fraction $f_i$ of each $b$ hadron and therefore the available statistics. While the probability $f_i$ of a $b\bar{b}$ pair hadronising into an $\Lambda_b^0$ baryon is about $f_{\Lambda_b^0}\sim18\%$ \protect\cite{fraction}, for each $\Xi_b$ baryon the production fraction is only approximately $f_{\Xi_b}\sim1.5\%$ \protect\cite{fraction}. For the $\Omega_b^-$ baryon, the production fraction is assumed to be $f_{\Omega_b^-}\sim0.5\%$ \protect\cite{fraction}, but this assumption is based on a measurement by the CDF experiment \protect\cite{CDF1} with only 16 $\Omega_{b}^{-}$ events. The low expected production probability makes the $\Omega_b^-$ difficult to study with respect to other weakly decaying $b$ baryons.\\
In order to search for effects beyond the SM, the properties of the $b$ baryons have to 
be known with high precision. The production fractions of these baryons need to be known to measure the absolute branching fractions of their decays. As a first step, the relative production fraction of the $\Omega_b^-$ baryon has been determined relatively to the $\Xi_b^-$ baryon for the first time at the Large Hadron Collider (LHC) by the LHCb experiment:
\begin{equation}
       \label{eqn:R}
       R = \frac{f_{\Omega_b^-}}{f_{\Xi_b^-}}\times\frac{\mathcal{B}(\Omega_b^-\to J/\psi\Omega^-)}{\mathcal{B}(\Xi_b^-\to J/\psi\Xi^-)}.
\end{equation}
These decay modes were chosen as they both provide a decay chain of weakly decaying particles that cannot be mimicked by any mesonic decays, leading to very low background contamination. The two final states of the $\Omega_b^-$ and $\Xi_b^-$ decay only differ by one particle since the $\Omega^-$ and the $\Xi^-$ are reconstructed in their decays to, \mbox{$\Lambda K^-)$} and ${\Lambda \pi^-}$, respectively. Given this particular topology, many systematic uncertainties cancel in the relative production fraction ratio $R$ (Eq. \protect\ref{eqn:R}). In addition, both decays have rather low energy releases making them 
excellent candidates to measure the $\Omega_b^-$ baryon mass through the mass difference with the $\Xi_b^-$ baryon. Since the 
$\Omega^-$ baryon has shorter lifetime compared to that of the $\Xi^-$, more $\Omega_b^-$ decays happen within the LHCb acceptance.
Therefore, it has a slightly better detection efficiency compared to the $\Xi_b^-$ baryon. \\
The measurement \protect\cite{paper} to determine the relative production fraction ratio $R$ and the $\Omega_b^-$ mass has been performed with a fully cut-based selection in a fiducial phase-space.

\section{The \texorpdfstring{$\Omega_{b}^{-}$}{Ωb⁻} mass measurement}
\begin{figure}[ht]
\begin{minipage}{0.5\linewidth}
\centerline{\includegraphics[height=2.25in]{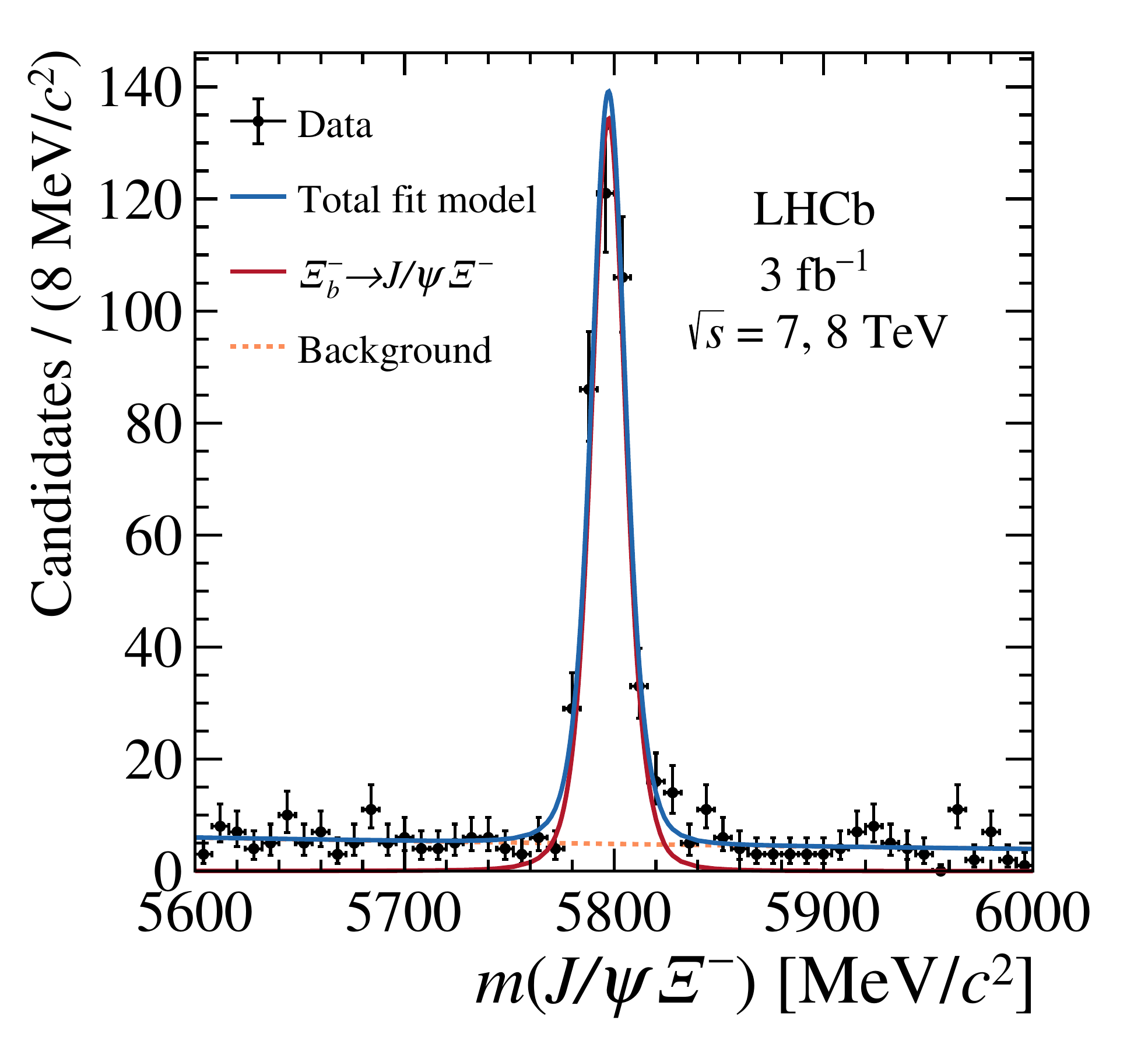}}
\end{minipage}
\begin{minipage}{0.5\linewidth}
\centerline{\includegraphics[height=2.25in]{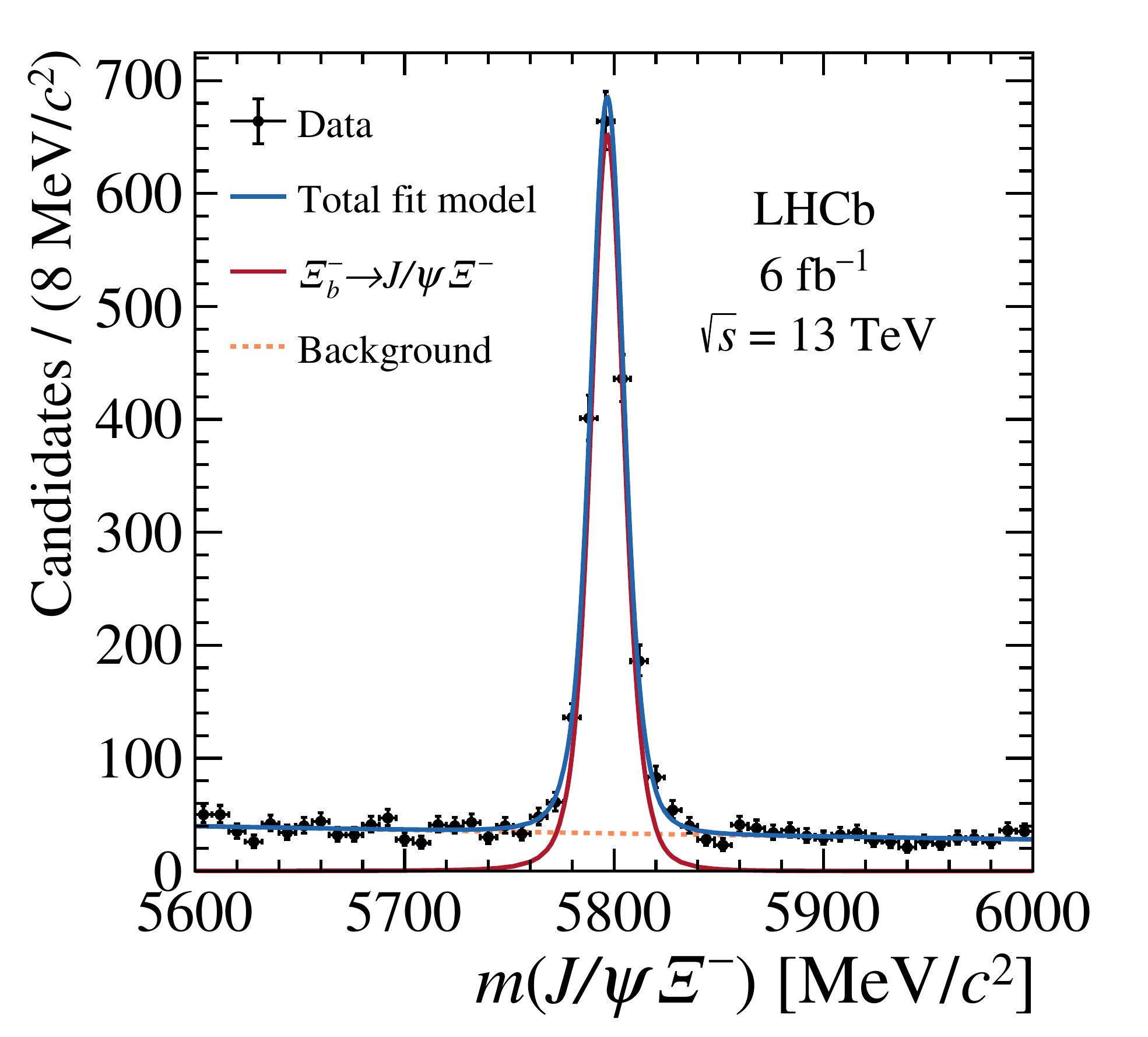}}
\end{minipage}

\begin{minipage}{0.5\linewidth}
\centerline{\includegraphics[height=2.25in]{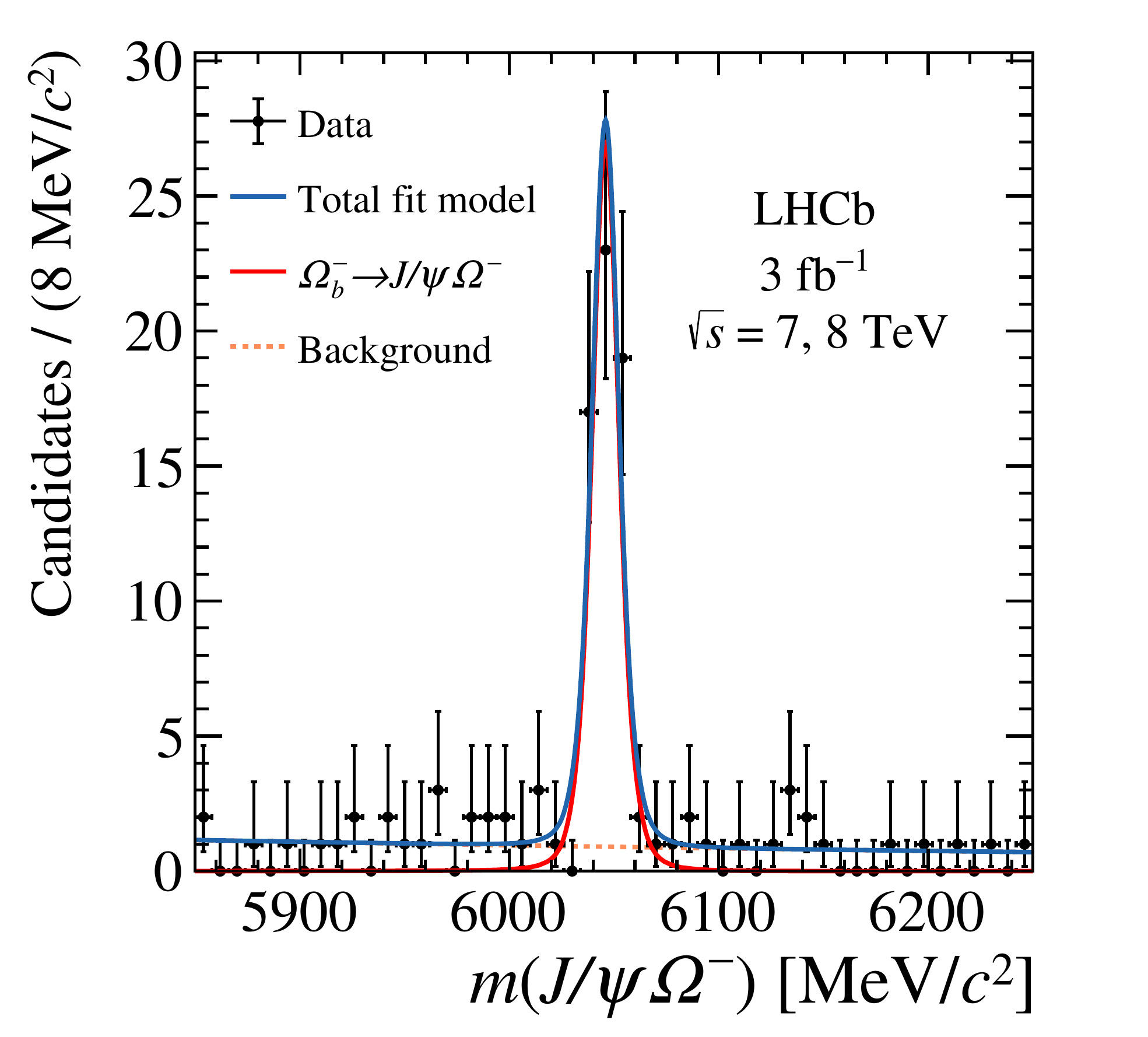}}
\end{minipage}
\begin{minipage}{0.5\linewidth}
\centerline{\includegraphics[height=2.25in]{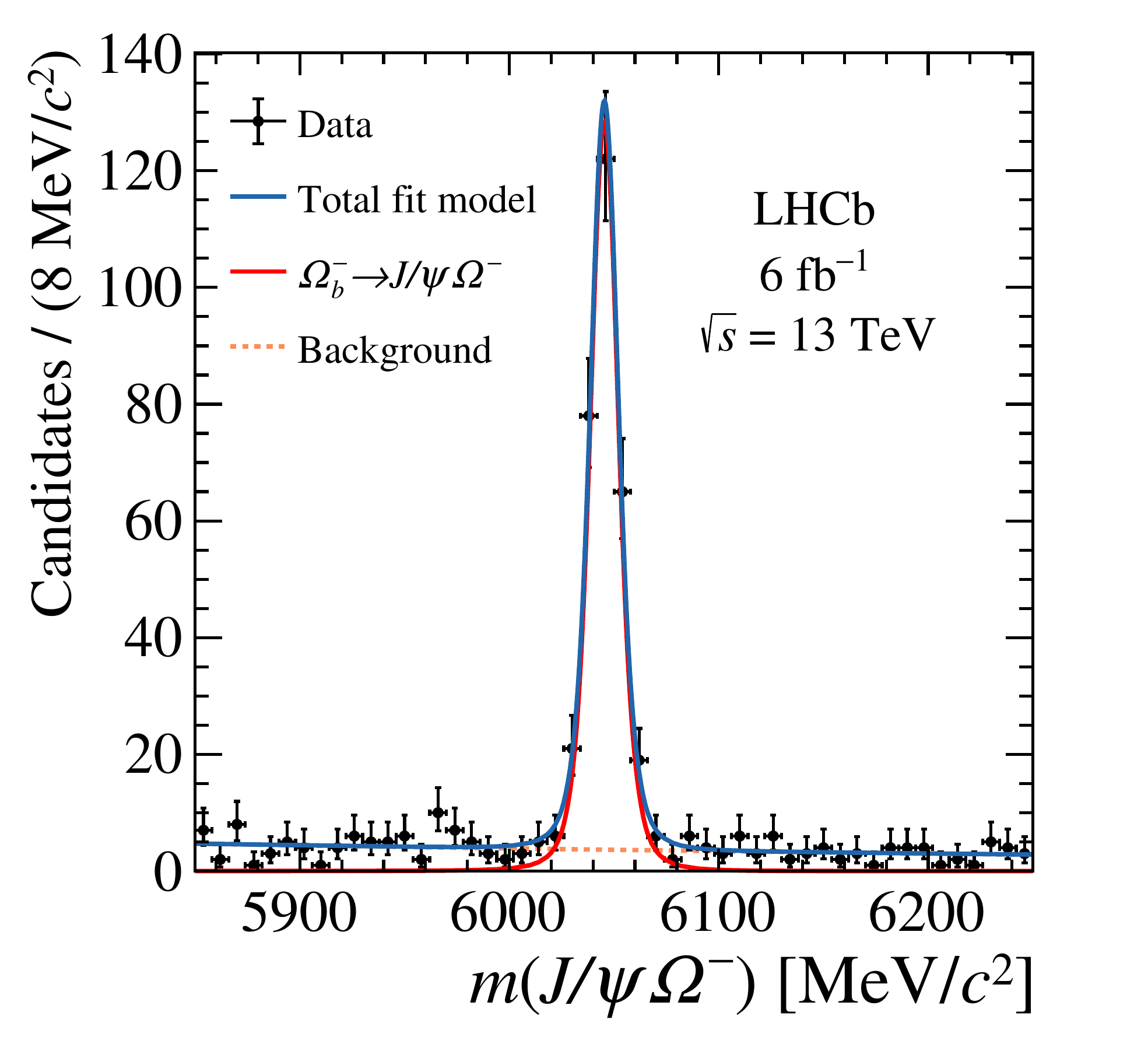}}
\end{minipage}x
\caption{Invariant-mass distributions in the (top) $\Xi_b^-\to  J/\psi\Xi^-$ and (bottom) $\Omega_b^-\to J/\psi\Omega^-$ datasets, in (left) Run 1 and (right) Run 2 data-taking periods used for the mass-difference measurement. (Taken from Ref.\protect\cite{paper})}
\label{fig:mOb}
\end{figure}

The full Run 1 and Run 2 LHCb dataset corresponding to an integrated luminosity of $9\;\mathrm{fb}^{-1}$ is used to determine the $\Omega_b^-$ baryon mass.
Instead of directly extracting the mass, the mass difference ${m(\Omega_b^-)-m(\Xi_b^-)}$ is measured. 
This choice allows to cancel the dominant systematic uncertainty due to the knowledge of the absolute momentum scale. The mass difference is kept as a free parameter and extracted from a simultaneous likelihood fit shown in Fig. \protect\ref{fig:mOb}. \\
\\
The resulting value is 
\begin{equation}
       \label{eqn:md}
       m(\Omega_b^-)-m(\Xi_b^-) = 248.54\pm0.51(\mathrm{stat})\pm 0.38(\mathrm{syst})\;\mathrm{MeV}/c^2.
\end{equation}
The dominant source for the systematic uncertainty is related to the poor knowledge of the hyperon masses. Together with the most precise determination of the $\Xi_b^-$ baryon mass \protect\cite{mxib} using the decay $\Xi_b^-\to\Xi_c^0\pi^-$ from the LHCb experiment, with a value of $m(\Xi_{b}^{-}) = 5797.33 \pm 0.24 \;(\mathrm{stat.})\pm 0.29 \;(\mathrm{syst.}) \;\mathrm{MeV}/c^2$, the $\Omega_{b}^{-}$ mass is determined to be 

\begin{equation}
       \label{eqn:mOb}
       m(\Omega_b^-)= 6045.9 \pm 0.5\;(\mathrm{stat.}) \pm 0.6\;(\mathrm{syst.}) \;\mathrm{MeV}/c^2.
\end{equation}
This measurement provides an improvement of more than a factor of two in precision with respect to the previous best measurement \protect\cite{mOb}.

\section{The relative production fraction \texorpdfstring{$R$}{R}}
\begin{figure}[ht]
\begin{minipage}{0.5\linewidth}
\centerline{\includegraphics[height=2.25in]{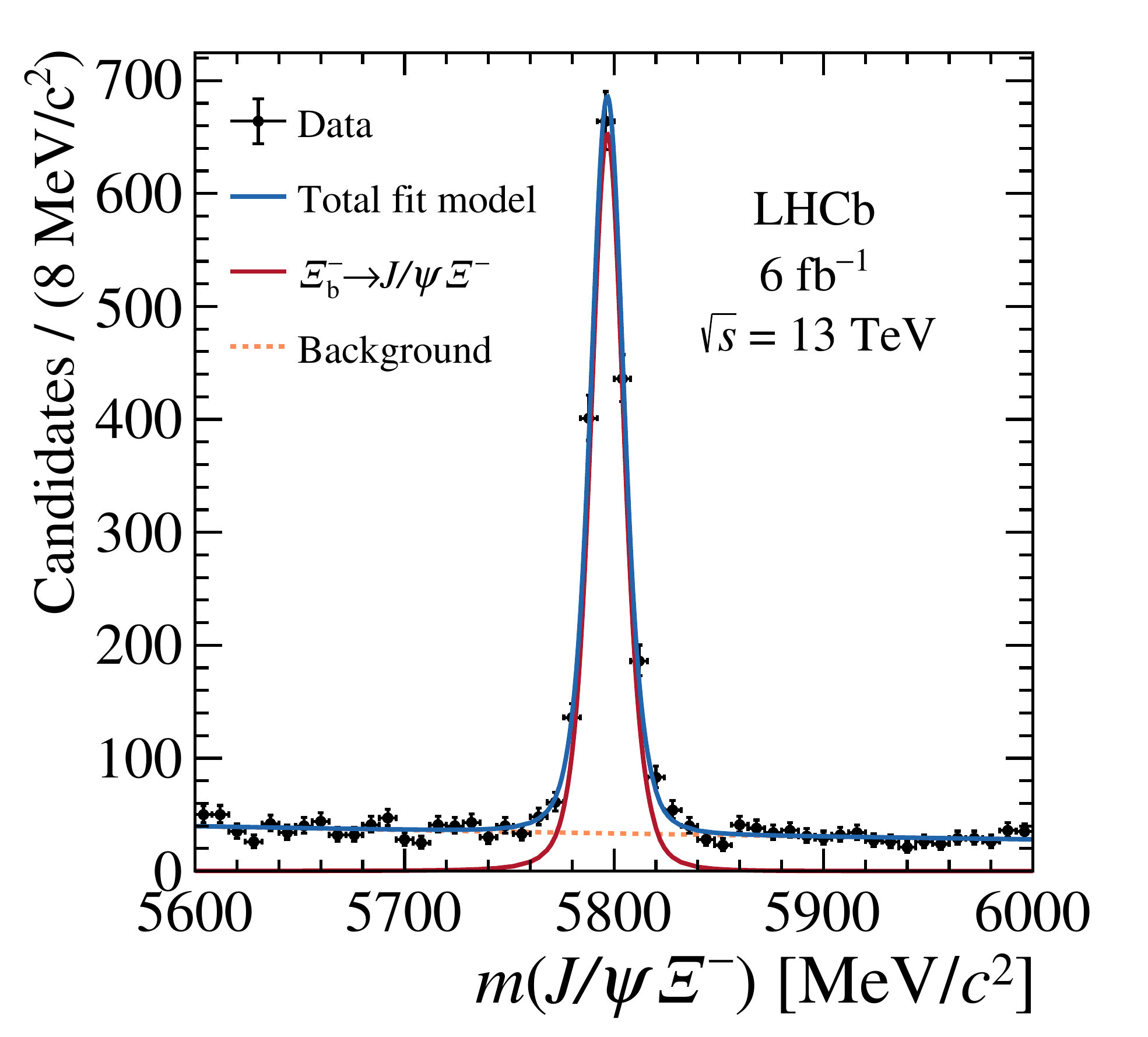}}
\end{minipage}
\begin{minipage}{0.5\linewidth}
\centerline{\includegraphics[height=2.25in]{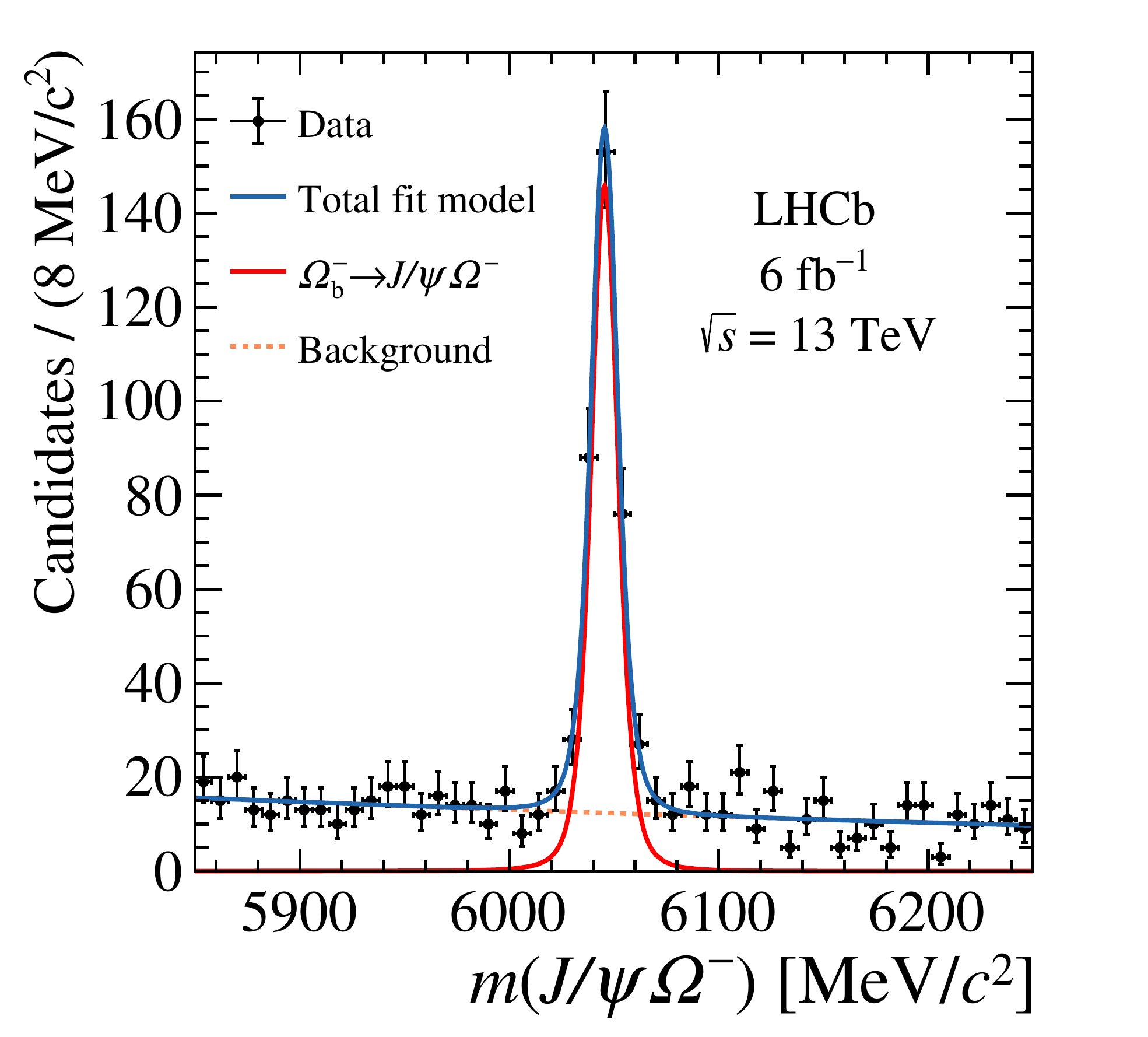}}
\end{minipage}
\caption{Invariant-mass distributions in the (left) $\Xi_b^-\to  J/\psi\Xi^-$ and (right) $\Omega_b^-\to J/\psi\Omega^-$ \mbox{Run 2} datasets used for the relative production estimation. (Taken from Ref.\protect\cite{paper})}
\label{fig:prod}
\end{figure}

Due to the energy dependence of the production fractions and the limited statistical power of the Run 1 dataset, the relative production fraction $R$ (see Eq. \protect\ref{eqn:R}) is only determined through a simultaneous likelihood fit to the Run 2 dataset. This sample corresponds to an integrated luminosity of $6\;\mathrm{fb}^{-1}$ recorded at a centre-of-mass energy of $13\;\mathrm{TeV}$. The selection is optimised to improve the sensitivity for the production fraction, and the corresponding mass fits are shown in Fig. \protect\ref{fig:prod}.

The relative production fraction is kept as a free fit parameter and measured to be
\begin{equation}
       R = 0.120\pm0.008(\mathrm{stat})\pm0.008(\mathrm{syst}).
\end{equation}
The dominant systematic uncertainties are due to simulation calibration linked to the small size of the $\Omega_b^-$ dataset and the poor knowledge of the $b$ baryon lifetimes. The measurement is in agreement within $1.2\sigma$ with the CDF measurement \protect\cite{CDF2} but the LHCb measurement suggests a larger relative production. In addition to the statistical limitation of the CDF measurement with $16$ $\Omega_b^-$ events, the larger nominal value can be related to the different production environments at the Tevatron and the LHC, leading to differing production cross-sections.

\section{Conclusion}
\begin{figure}[ht]
\begin{minipage}{\linewidth}
\centerline{\includegraphics[width=0.5\linewidth]{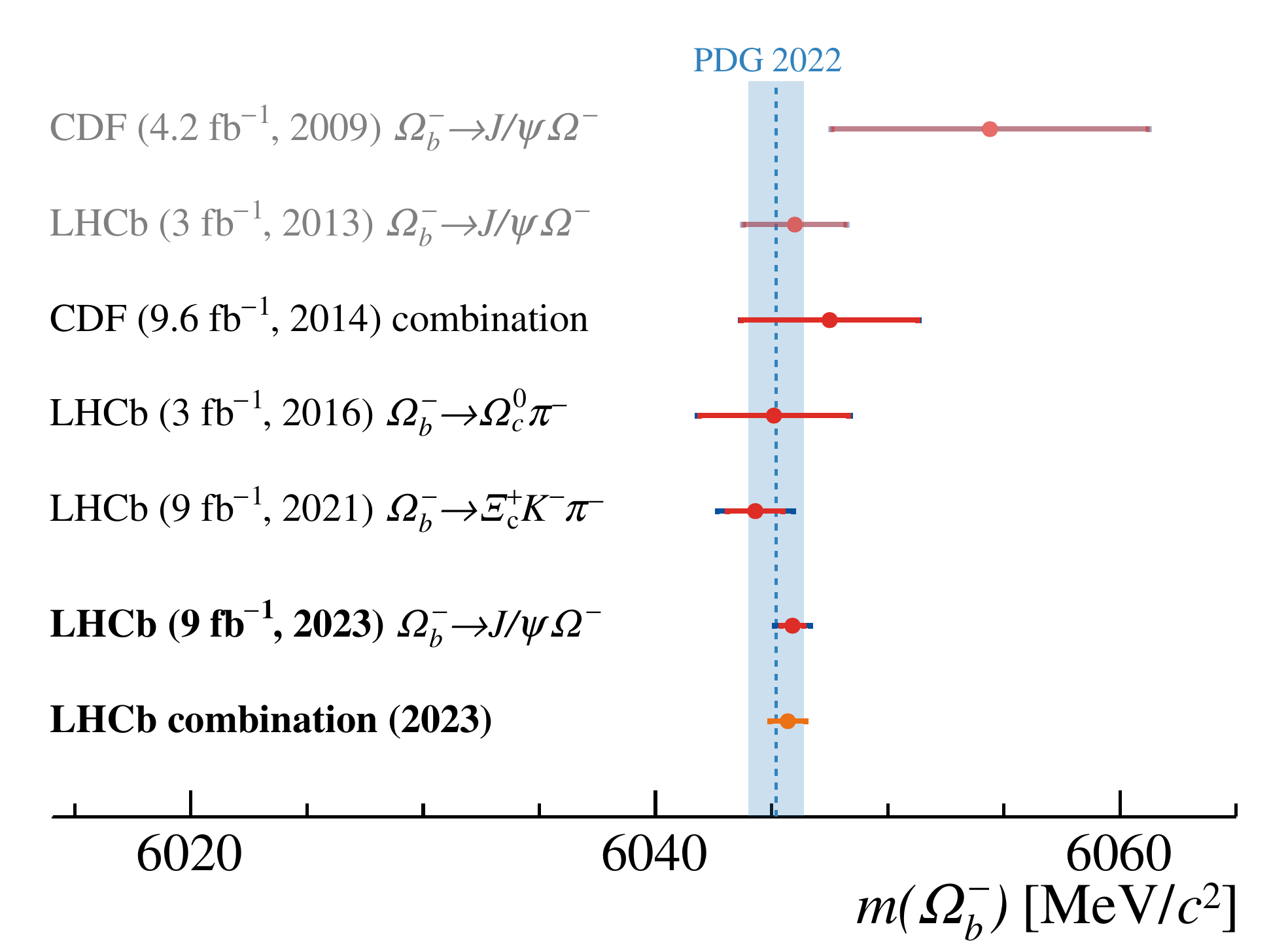}}
\end{minipage}
\caption{Overview of the $\Omega_{b}^{-}$ mass measurements to date (red points)~\protect\cite{CDF1,mOb,CDF2,mOb1,mOb2}, with corresponding statistical and systematic uncertainties marked in red and blue, respectively, as well as the current world average by the Particle Data Group~\protect\cite{PDG2022} (light-blue band). The superseded measurements are shown with lighter colours. The new LHCb average is presented in orange, showing the total uncertainty. (Taken from Ref. \protect\cite{paper})}
\label{fig:overview}
\end{figure}

These proceedings present the most precise measurement of the $\Omega_b^-$ baryon mass using the LHCb 
dataset corresponding to an integrated luminosity of $9\;\mathrm{fb}^{-1}$. The measurement agrees with the world average and previous 
measurements, which are displayed in Fig. \protect\ref{fig:overview}. 

In addition, the first determination of the relative production 
fraction of the $\Omega_b^-$ baryon at the LHC at a centre-of-mass energy of $\sqrt{s}=13\;\mathrm{TeV}$ was presented
\begin{equation}
       \frac{f_{\Omega_b^-}}{f_{\Xi_b^-}}\times\frac{\mathcal{B}(\Omega_b^-\to\Omega_b^- J/\psi)}{\mathcal{B}(\Xi_b^-\to\Xi_b^- J/\psi)} = 0.120\pm0.008(\mathrm{stat.})\pm0.008(\mathrm{syst.}),
\end{equation}
using a dataset of $6\;\mathrm{fb}^{-1}$.
Further theoretical input on the branching fractions is needed to disentangle the ratio of production fractions from the ratio of branching fractions and enable to perform a first absolute branching fraction measurement of $\Omega_b^-$ baryon decays. 
\section*{Acknowledgments}
J.N. acknowledges support from the European Research Council (ERC) under the European Union's Horizon 2020 research and 
innovation programme under grant agreement No. 714536: PRECISION, the German Federal Ministry of Education and Research 
(BMBF, grant no. 05H21PECL1) within ErUM-FSP T04, and the German Academic Scholarship Foundation.

\end{document}